\documentclass[12pt,draftcls,onecolumn]{IEEEtran}

\def \be {\begin{equation}}

\def \ee {\end{equation}}

\def \nn {\nonumber}

\usepackage[dvips]{graphicx}

\usepackage[cmex10]{amsmath}
\usepackage{amssymb}

\begin{document}

\title{Similarity Decomposition Approach to Oscillatory Synchronization for Multiple Mechanical Systems With a Virtual Leader}

\author{Hanlei Wang
\thanks{The author is with Science and Technology on Space Intelligent Control Laboratory, Beijing Institute of
Control Engineering, Beijing 100190, China (e-mail: hlwang.bice@gmail.com).}}


\maketitle

\begin{abstract}
This paper addresses the oscillatory synchronization problem for multiple uncertain mechanical systems with a virtual leader, and the interaction topology among them is assumed to contain a directed spanning tree. We propose an adaptive control scheme to achieve the goal of oscillatory synchronization. Using the similarity decomposition approach, we show that the position and velocity synchronization errors between each mechanical system (or follower) and the virtual leader converge to zero. The performance of the proposed adaptive scheme is shown by numerical simulation results.
\end{abstract}



\section{Introduction}

Synchronization problem for multi-agent systems has been intensively studied in recent years \cite{Jadbabaie2003_TAC,Olfati-Saber2004_TAC,Ren2005_TAC,Jadbabaie2004_ACC,Chopra2006,Ren2007_RNC,Lee2007_TAC,Ren2008_Aut,Chopra2009_TAC,Su2009_Aut,Cheah2009_Aut,Wang2013cyclic_ASME}, due to the universal existence of synchronized phenomena in nature and also its close relation to many engineering applications. A common practice in the current literature is to adopt the neighboring-information-based control action so that some kind of group behavior is attained (see, e.g., \cite{Olfati-Saber2004_TAC,Ren2008_Book}).

There are roughly two branches of research on synchronization problem (see, e.g., \cite{Ren2008_Aut}). The first branch focuses on the consensus problem and the second concentrates on the oscillator synchronization problem. The work in \cite{Jadbabaie2003_TAC,Olfati-Saber2004_TAC,Ren2005_TAC,Chopra2006,Ren2007_RNC,Lee2007_TAC,Nuno2011_TAC,Wang2013b_Aut,Wang2013_TAC}
can be categorized into the first branch, whose main goal is to synchronize the states of interest of the agents to a common value (in many cases, it is constant). The research in \cite{Jadbabaie2004_ACC,Paley2007_CSM,Ren2008_Aut,Chopra2009_TAC,Su2009_Aut} belongs to the second branch, and the control objective in these studies, different from the one in the first branch, is to achieve certain oscillatory synchronized motion (i.e., the equilibrium, in most cases, is oscillatory). Another interesting result appears in \cite{Chopra2006}, which discusses the oscillatory synchronization of multiple pendulums. Other studies (see, e.g., \cite{Yu2010_SMC,Liu2013_SCL,Su2011_Aut,Li2013_TAC}) also present control schemes that can realize certain kind of oscillatory synchronization due to the explicit inclusion/presence of Lipschitz nonlinearity in the closed-loop network dynamics.

Most of the above results presented in the second branch, however, are confined to agents with exactly known linear dynamics. For example, the results in \cite{Ren2008_Aut,Su2009_Aut} rely on the assumption that the mass agents are identical, and in the case of nonidentical mass agents, this assumption  shall be equivalent to the requirement that their masses be precisely known. The pendulum-model-based synchronization scheme in \cite{Chopra2006} does not need the accurate knowledge of the masses, yet, it requires the lengths of the pendulums to be exactly the same. From a control viewpoint, if we expect to achieve oscillatory synchronized motion like the one generated by the networked pendulums in \cite{Chopra2006}, the model of the mass agent, again, must be known accurately. The case is similar for the scheme relying on Lipschitz nonlinearity (e.g., \cite{Yu2010_SMC,Liu2013_SCL,Su2011_Aut,Li2013_TAC}). It is emphasized that, here, the Lipschitz nonlinearity and the nonlinear sine function in \cite{Chopra2006} are not considered to be the model nonlinearity, but simply as a way to realize oscillatory motion.


However, in many practical applications, it is unrealistic to assume that the dynamic models of the agents are linear and exactly known, e.g., robot manipulators, spacecraft, mobile robots, and unmanned aerial vehicles (UAVs). The dynamics governing these agents, also known as \emph{mechanical dynamics} or \emph{Euler-Lagrange dynamics}, is not only highly nonlinear but often uncertain (see, e.g., \cite{Slotine1991_Book}). Some attempts along this direction are the leader-follower control schemes in \cite{Nuno2011_TAC,Mei2012_Aut,Cai2013_CCC}. The control scheme in \cite{Nuno2011_TAC} is not distributed since it requires that the virtual leader's information be known by each follower, and those in \cite{Mei2012_Aut,Cai2013_CCC} are indeed distributed thanks to the employment of distributed observers (oscillatory synchronization can certainly be achieved by properly designing
an oscillatory motion for the virtual leader). The use of these distributed observers (which require the communication of the observed signals among the followers), unfortunately, complicates the control schemes. 
The control schemes in \cite{Ding2013_TAC,Ding2013_CDC}, without using any distributed observers, achieve oscillatory synchronization for general nonlinear agents with a virtual leader. Nevertheless, the result given in \cite{Ding2013_TAC} requires the exact knowledge of the dynamic models of the agents. This restrictive assumption is relaxed in \cite{Ding2013_CDC}, which takes into account second-order agents with matched uncertainties on undirected graphs, yet, the extension of \cite{Ding2013_CDC} to the more general directed graphs is still unclear since it is well known that the graph Laplacian in that case is usually asymmetrical.

In this paper, we propose an adaptive control scheme to realize oscillatory synchronization for a network of multiple uncertain mechanical systems and a virtual leader, and the proposed scheme does not employ any distributed observers, in contrast to \cite{Mei2012_Aut,Cai2013_CCC}. Relying on the similarity decomposition approach \cite{Wang2013b_Aut} and using Lyapunov-like analysis and input-output analysis, we show that the position and velocity synchronization errors between each mechanical system and the virtual leader with oscillatory motion converge to zero. Unlike the results in \cite{Ren2008_Aut,Chopra2006} that require the exact knowledge of the mass properties of the linear agents if their masses are nonidentical (very common in practice), our result, due to the use of adaptive strategy, no longer relies on this relatively restrictive assumption and additionally considers the systems that are governed by nonlinear dynamics rather than linear dynamics. In this sense, our result extends \cite{Ren2008_Aut,Chopra2006} to the case of nonidentical mechanical systems with high nonlinearity and parametric uncertainty. Although many researchers have made some extensions from linear systems to systems with Lipschitz nonlinearity (e.g., \cite{Yu2010_SMC,Liu2013_SCL,Su2011_Aut,Li2013_TAC}), this kind of nonlinearity is rather weak as compared with the nonlinear terms in mechanical systems (which generally include the squares and mutual multiplications of the derivatives of the generalized coordinates). In addition, our result allows the interaction topology to be directed and is thus more general than the undirected topology case considered in \cite{Ding2013_CDC}. Other control schemes that are possibly related to the one in the present work are consensus/flocking control schemes for multiple mechanical systems without a leader in \cite{Nuno2011_TAC,Min2012_SCL,Mei2012_Aut,Wang2013b_Aut,Wang2013_TAC,Wang2014_TAC}. The position consensus equilibrium in \cite{Nuno2011_TAC,Min2012_SCL,Mei2012_Aut} is unknown (possibly unbounded), and the position consensus equilibrium in \cite{Wang2013_TAC,Wang2014_TAC} and the velocity consensus equilibrium in \cite{Wang2013b_Aut} are constant. The control scheme presented here, however, ensures oscillatory coordination of multiple mechanical systems with a virtual leader, i.e., the position and velocity consensus equilibria are both oscillatory.

\section{Preliminaries}

\subsection{Graph Theory}

Let us briefly introduce the theory of directed graphs based on \cite{Godsil2001_Book,Olfati-Saber2004_TAC,Ren2005_TAC,Ren2008_Book}. We take into account $n$ mechanical systems (also called followers) with a virtual leader, and for the convenience of later reference, we attach index 0 to the virtual leader, and index $i$ to the $i$-th follower, $i=1,2,\dots,n$. As is now commonly done, we utilize a directed graph ${\mathcal G}^\ast=\left({\mathcal V}^\ast,{\mathcal E}^\ast\right)$ to describe the interaction topology among the virtual leader and the followers, where $\mathcal{V}^\ast=\left\{0,1,\dots,n\right\}$ is the vertex set that denotes the collection of the virtual leader and all the followers, and $\mathcal{E}^\ast\subseteq \mathcal{V}^\ast\times\mathcal{V}^\ast$ is the edge set that describes the information flow among the virtual leader and the followers. The set of neighbors of the $i$-th follower is denoted by $\mathcal{N}_i^\ast=\left\{j|(i,j)\in\mathcal{E}^\ast\right\}$, and the set of neighbors of the virtual leader is denoted by ${\mathcal N}_0^\ast$, which is obviously an empty set. A directed graph is said to contain a spanning tree if there is a vertex $k^\ast\in {\mathcal V}^\ast$ such that any other vertex of the graph has a directed path to vertex $k^\ast$. The weighted adjacency matrix $\mathcal{W}^\ast=\left[w_{ij}\right]$ associated with $\mathcal G^\ast$ is defined according to the rule that $w_{ij}>0$ if $ j\in\mathcal{N}_i^\ast$, and $w_{ij}=0$ otherwise, $\forall i,j=0,1,\dots,n$. The Laplacian matrix $\mathcal{L}_w^\ast=\left[\ell_{w,ij}\right]$ associated with $\mathcal G^\ast$ is defined as $\ell_{w,ij}=\Sigma_{k=0}^n w_{ik}$ if $i=j$, and $\ell_{w,ij}=-w_{ij}$ otherwise, $\forall i,j=0,1,\dots,n$. Some basic properties of the Laplacian matrix $\mathcal{L}_w^\ast$ are described by the following lemma.

\emph{Lemma 1 (\cite{Ren2005_TAC,Ren2008_Book}):} If the graph $\mathcal G^\ast$ contains a spanning tree rooted at vertex $0$, then
 \begin{enumerate}
 \item
  $\mathcal{L}_w^\ast$ has a simple zero eigenvalue and all the other eigenvalues of $\mathcal{L}_w^\ast$ are in the open right half plane (RHP);

  \item the vectors $\gamma^\ast=\left[1,0,\dots,0\right]^T $ and $1_{n+1}=\left[1,1,\dots,1\right]^T$ are the left and right eigenvectors of $\mathcal{L}_w^\ast$ associated with its zero eigenvalue, respectively, i.e., $\gamma^{\ast T} \mathcal{L}_w^\ast=0$ and $\mathcal{L}_w ^\ast 1_{n+1}=0$;

  \item $\text{rank}\left(\mathcal{L}_w^\ast\right)=n$.
\end{enumerate}

\subsection{Equations of Motion of Mechanical Systems}

The equations of motion of the $i$-th mechanical system (i.e., the $i$-th follower) can be written as \cite{Spong2006_Book,Slotine1991_Book}
\be
\label{eq1}
M_i(q_i)\ddot{q}_i+C_i(q_i,\dot{q}_i)\dot{q}_i+g_i(q_i)=\tau_i
\ee
where $q_i \in R^m$ is the generalized position (or configuration), $M_i \left( {q_i }
\right) \in R^{m\times m}$ is the inertia matrix, $C_i \left( {q_i ,\dot
{q}_i } \right) \in R^{m\times m}$ is the Coriolis and centrifugal matrix,
$g_i \left( {q_i } \right) \in R^m$ is the gravitational torque, and
$\tau _i \in R^m$ is the exerted control torque.

Three familiar properties associated with the dynamic model (\ref{eq1}) that shall be useful for the subsequent controller design and stability analysis are listed as follows (see, e.g., \cite{Spong2006_Book,Slotine1991_Book}).

\emph{Property 1:} The inertia matrix $M_i(q_i)$ is symmetric and uniformly positive definite.

\emph{Property 2:} The Coriolis and centrifugal matrix $C_i(q_i,\dot{q}_i)$ can be appropriately chosen such that $\dot{M}_i(q_i)-2C_i(q_i,\dot{q}_i)$ is skew-symmetric.

\emph{Property 3:} The dynamic model (\ref{eq1}) depends linearly on a constant parameter vector $a_i$, thus yielding
\be
M_i(q_i)\dot{\zeta}+C_i(q_i,\dot{q}_i)\zeta+g_i(q_i)=Y_i(q_i,\dot{q}_i,\zeta,\dot{\zeta})a_i
\ee
where $Y_i(q_i,\dot{q}_i,\zeta,\dot{\zeta})$ is the regressor matrix, $\zeta\in R^m$ is a differentiable vector, and $\dot{\zeta}$ is the time derivative of $\zeta$.

\section{Adaptive Oscillatory Synchronization}

In this section, we will seek an adaptive control scheme to realize the oscillatory synchronization of the $n$ mechanical systems (followers) with a virtual leader. The virtual leader considered here is the same as the one in \cite{Ren2008_Aut}, whose behavior can be described by the following standard oscillatory dynamics
\be
\label{eq3}
\ddot{q}_0=-\alpha q_0
\ee
where $\alpha>0$ is a design constant, and $q_0\in R^m$ denotes the position of the virtual leader. Then, the control objective is to drive the state (i.e., position and velocity) of each follower to the oscillatory state generated by (\ref{eq3}).

For the $i$-th follower, we define a new reference velocity of the following form
\be
\label{eq4}
\dot{q}_{r,i}=-\Sigma_{j\in\mathcal{N}_i^\ast}w_{ij}\left(q_i-q_j\right)\underbrace{-\alpha\int_0^t q_i(r)dr}_\text{integral action}
\ee
where the \emph{integral action} will be shown to be essential for realizing the oscillatory synchronization.
Differentiating equation (\ref{eq4}) with respect to time yields the reference acceleration
\be
\label{eq5}
\ddot{q}_{r,i}=-\Sigma_{j\in\mathcal{N}_i^\ast}w_{ij}\left(\dot q_i-\dot q_j\right)-\alpha q_i.
\ee

Then, let us define a sliding vector as
\be
\label{eq6}
s_i=\dot{q}_i-\dot{q}_{r,i}
\ee
whose derivative with respect to time can be written as
\be
\label{eq7}
\dot{s}_i=\ddot{q}_i+\Sigma_{j\in\mathcal{N}_i^\ast}w_{ij}\left(\dot q_i-\dot q_j\right)+\alpha q_i.
\ee

We propose the following control law for the $i$-th follower
\be
\label{eq8}
\tau_i=-K_i s_i+Y_i(q_i,\dot{q}_i,\dot{q}_{r,i},\ddot{q}_{r,i})\hat{a}_i
\ee
where $K_i$ is a symmetric positive definite matrix, and $\hat{a}_i$ is the estimate of the parameter $a_i$, which is updated by the adaptation law
\be
\label{eq9}
\dot{\hat{a}}_i=-\Gamma_i Y_i^T(q_i,\dot{q}_i,\dot{q}_{r,i},\ddot{q}_{r,i})s_i
\ee
where $\Gamma_i$ is a symmetric positive definite matrix.

\emph{Remark 1:} The adaptive controller given by (\ref{eq8}) and (\ref{eq9}) is basically the same as the well-known Slotine and Li adaptive control \cite{Slotine1987_IJRR}, and the difference lies in the definition of the new reference velocity and acceleration which incorporate the neighboring information of each follower.

Substituting the control law (\ref{eq8}) into the dynamics (\ref{eq1}) yields
\be
\label{eq10}
M_i(q_i)\dot{s}_i+C_i(q_i,\dot{q}_i)s_i=-K_i s_i+Y_i(q_i,\dot{q}_i,\dot{q}_{r,i},\ddot{q}_{r,i})\Delta a_i
\ee
where $\Delta a_i=\hat{a}_i-a_i$ is the parameter estimation error.

The closed-loop behavior of the $i$-th follower can then be described by 
\be
\label{eq11}
\begin{cases}
\dot{q}_0=-\alpha\int_0^t q_0(r)dr+\dot{q}_0(0),\\
\dot{q}_i=-\Sigma_{j\in\mathcal{N}_i^\ast}w_{ij}\left(q_i-q_j\right)-\alpha\int_0^t q_i(r)dr+s_i,\\
M_i(q_i)\dot{s}_i+C_i(q_i,\dot{q}_i)s_i\\
=-K_i s_i+Y_i(q_i,\dot{q}_i,\dot{q}_{r,i},\ddot{q}_{r,i})\Delta a_i, \\
\dot{\hat{a}}_i=-\Gamma_i Y_i^T(q_i,\dot{q}_i,\dot{q}_{r,i},\ddot{q}_{r,i})s_i
\end{cases}
\ee
where the integration of (\ref{eq3}) is included since the state of the virtual leader directly/indirectly influences that of each follower.

Stacking up the first subsystem and all the subsystems expressed as the second one in (\ref{eq11}) yields
\begin{align}
\label{eq12}
\overbrace{\dot{q}^\ast=-\left(\mathcal{L}_w^\ast\otimes I_m\right) q^\ast-\alpha \int_0^t q^\ast(r)dr+d_0^\ast}^{\Psi}
+s^\ast
\end{align}
where $\otimes$ denotes the standard Kronecker product \cite{Brewer1978_TCS}, $q^\ast=\left[q_0^T,q_1^T,\dots,q_n^T\right]^T$, $d_0^\ast=\left[\dot{q}_0^T(0),0_{mn}^T\right]^T$ is a constant vector,  $s^\ast=\left[0_m^T,s_1^T, s_2^T,\dots,s_n^T\right]^T$, and $I_m$ is the $m\times m$ identity matrix.

Although the work reported in \cite{Ren2008_Aut} has already presented the stability and convergence properties of the differentiated form of the system $\Psi$, i.e., $$\underbrace{\ddot{q}^\ast=-\left(\mathcal{L}_w^\ast\otimes I_m\right)\dot{q}^\ast-\alpha q^\ast}_{{d\Psi}/{dt}},$$
it is not so clear about the properties of the outputs $q_i$, $i=1,2,\dots,n$  of (\ref{eq12}) due to the presence of the external input signal $s^\ast$.

Let us adopt the similarity decomposition in \cite{Wang2013b_Aut} to analyze the system (\ref{eq12}), which relies on the following coordinate transformation \cite{Lee2007_TAC,Li2010_TAC}
\be
\label{eq13}
\xi=\left(T\otimes I_m\right) q^\ast
\ee
where the transformation matrix $T\in R^{(n+1)\times(n+1)}$ is
\be
T =\begin{bmatrix}
 {1 }  & {0 }  & {0 }  & \cdots
 & {0 }  \\
 1  & { - 1}  & 0  & \cdots  & 0  \\
 0  & 1  & { - 1}  & \cdots  & 0  \\
 \vdots  & \vdots  & \ddots  & \ddots  & \vdots
 \\
 0  & 0  & \cdots  & 1  & { - 1} \end{bmatrix}
\ee
and the vector $\xi=\left[\xi_1^T,\xi_E^T\right]^T$, in which $\xi_1=q_0$ and $\xi_E=\left[q_0^T-q_1^T,q_1^T-q_2^T,\dots,q_{n-1}^T-q_n^T\right]^T$. Applying the similarity transformation based on the transformation equation (\ref{eq13}) (following \cite{Wang2013b_Aut}) to (\ref{eq12}) yields
\begin{align}
\label{eq15}
\dot{\xi}=&-\left[\left(T \mathcal{L}_w^\ast T^{-1}\right)\otimes I_m\right]\xi-\alpha\int_0^t \xi(r)dr\nn\\&+ \left(T\otimes I_m\right)\left(s^\ast+d_0^\ast\right)
\end{align}
where the matrix $T \mathcal{L}_w^\ast T^{-1}$ can be decomposed as \cite{Wang2013b_Aut}
\be
\label{eq16}
T\mathcal{L}_w^\ast T^{-1}=\text{diag}\left[0,\bar{L}_w^\ast\right]
\ee
where the matrix $\bar{L}_w^\ast\in R^{n\times n}$ satisfies the property that all its eigenvalues are in the open RHP if the interaction graph among the virtual leader and the $n$ followers contains a spanning tree.


\emph{Remark 2: } The Jordan form of a Laplacian matrix dates back to the result in \cite{Olfati-Saber2004_TAC} (concerning strongly connected directed graphs), and the extension to directed graphs only containing a spanning tree appears in, e.g., \cite{Ren2005}. Let ${\cal L}_w\in R^{p\times p}$ be the Laplacian matrix associated with a directed graph containing a spanning tree. The relationship between $\mathcal{L}_w$ and its Jordan form can be written as \cite{Olfati-Saber2004_TAC,Ren2005}
\be
\label{eq17a}
\mathcal{L}_w=D J D^{-1}
\ee
where the Jordan form $J=\text{diag}\left[0,\bar J\right]$ with $\bar{J}\in R^{(p-1)\times(p-1)}$ having the property that all its eigenvalues are in the open RHP, the first column of $D$ is $1_p=\left[1,1,\dots,1\right]^T$ and the first row of $D^{-1}$ is a nonnegative vector $\gamma=\left[\gamma_1,\gamma_2,\dots,\gamma_p\right]^T$ satisfying the property that $\gamma^T \mathcal{L}_w=0$ and $\Sigma_{k=1}^p\gamma_k=1$. The transformation (\ref{eq17a}) and the property of $\bar J$ are also exploited in \cite{Meng2011_SMC} to handle the consensus problem under input and communication delays. A more intuitive formulation of equation (\ref{eq17a}) can be obtained by letting $T_0=D^{-1}$ as
\be
\label{eq18a}
T_0 \mathcal{L}_w T_0^{-1}=\text{diag}\left[0,\bar{J}\right].
\ee
Due to \cite{Olfati-Saber2004_TAC,Ren2005}, $T_0$ can be written as
\be
T_0=\begin{bmatrix}
\gamma^T\\
\bar{T}_0
\end{bmatrix}
\ee
where $\bar T_0\in R^{(p-1)\times p}$ and $\text{rank}(\bar{T}_0)=p-1$, and in addition, $\bar{T}_0$ obviously satisfies the following property (since $T_0T_0^{-1}=I_p$, where $I_p$ is the $p\times p$ identity matrix)
\be
\label{eq20a}
\bar{T}_01_p=0.
\ee
The transformation $T_1$ in \cite{Lee2007_TAC,Li2010_TAC} (other forms of $T_1$ can be found in, e.g., \cite{Lee2012_Aut}) is
\be
\label{eq21a}
T_1 =\begin{bmatrix}
 \gamma _1   & \gamma _2   & {\gamma _3 }  & \cdots
 & {\gamma _p }  \\
 1  & { - 1}  & 0  & \cdots  & 0  \\
 0  & 1  & { - 1}  & \cdots  & 0  \\
 \vdots  & \vdots  & \ddots  & \ddots  & \vdots
 \\
 0  & 0  & \cdots  & 1  & { - 1}
 \end{bmatrix}
\ee
which obviously satisfies property (\ref{eq20a}). However, in general, $T_1$ is different from $T_0$ and the similarity decomposition [e.g., equation (\ref{eq16})] in \cite{Wang2013b_Aut} is also unlike (\ref{eq17a}) in that, usually, it does not give rise to the Jordan form of $\mathcal{L}_w$.

Using the similarity decomposition (\ref{eq16}) and exploiting the standard constant disturbance compensation capability of the integral action, we can rewrite equation (\ref{eq15}) as
\be
\label{eq17}
\begin{cases}
\dot{q}_0=-\alpha\int_0^t q_0(r)dr+\dot{q}_0(0),\\
\overbrace{\dot{\xi}_E=-\left(\bar{L}_w^\ast\otimes I_m\right){\xi}_E-\alpha \left[\int_0^t \xi_E(r)dr-\alpha^{-1}d_0\right]}^{\Psi_r}\\+s_E^\ast
\end{cases}
\ee
where $s_E^\ast=\left[0_m^T-s_1^T, s_1^T-s_2^T, s_2^T-s_3^T,\dots,s_{n-1}^T-s_n^T\right]^T$ and $d_0=\left[\dot{q}_0^T(0),0_{m\left(n-1\right)}^T\right]^T$. The decomposition of (\ref{eq12}) into two subsystems given by (\ref{eq17}) is similar to \cite{Lee2007_TAC}, yet the decomposition technique used here is different from the one in \cite{Lee2007_TAC} (which is based on a congruence transformation). 

Let $\sigma_E=\int_0^t \xi_E(r)dr-\alpha^{-1}d_0$, and then, equation (\ref{eq17}) can be rewritten as
\be
\label{eq18}
\begin{cases}
\dot{q}_0=-\alpha\int_0^t q_0(r)dr+\dot{q}_0(0),\\
\overbrace{\ddot{\sigma}_E=-\left(\bar{L}_w^\ast\otimes I_m\right)\dot{\sigma}_E-\alpha\sigma_E}^{\Psi_r}+s_E^\ast
\end{cases}
\ee
where the property of the system $\Psi_r$ can be characterized by the following lemma.

\emph{Lemma 2:} If the interaction graph among the virtual leader and the $n$ followers contains a spanning tree rooted at vertex 0, then all the poles of $\Psi_r$ in (\ref{eq18}) are located in the open left half plane (LHP).

\emph{Proof: } The system (\ref{eq18}) reduces to the one considered in \cite{Ren2008_Aut} if the external input $s^\ast_E$ is ruled out, or more precisely, the system $\Psi$ in (\ref{eq12}). In the case that the interaction graph among the virtual leader and the $n$ followers contains a spanning tree rooted at vertex 0, it is demonstrated in \cite{Ren2008_Aut} that $\Psi$ contains two simple poles on the imaginary axis (the locations of which are determined by the parameter $\alpha$), and all the other poles of $\Psi$ are in the open LHP. Here, by the similarity decomposition, the two simple poles are both contained in the first subsystem of (\ref{eq18}). In fact, according to the standard linear system theory, the first subsystem in (\ref{eq18}) indeed include two poles on the imaginary axis, i.e., $\bar p_{1}=j^\ast \sqrt{\alpha}$ and $\bar p_{2}=-j^\ast\sqrt{\alpha}$, where $j^\ast=\sqrt{-1}$ denotes the imaginary unit.
Therefore, all the poles of the linear system $\Psi_r$ are those of $\Psi$ that are located in the open LHP. \hfill {\small $\blacksquare$}

We are presently ready to give the following theorem.

\emph{Theorem 1:} The control law (\ref{eq8}) and the parameter adaptation law (\ref{eq9}) ensure the convergence of the position and velocity synchronization errors between each follower and the virtual leader provided that the graph among the virtual leader and the $n$ followers contains a spanning tree rooted at vertex 0, i.e., $q_i(t)\to q_0(t)$ and $\dot{q}_i(t)\to \dot{q}_0(t)$ as $t\to\infty$, $\forall i=1,2,\dots,n$.

\emph{Proof: } Following \cite{Slotine1987_IJRR,Ortega1989_Aut}, we take into consideration the Lyapunov-like function candidate  $V_i=(1/2)s_i^T M_i(q_i)s_i+(1/2)\Delta a_i^T \Gamma_i^{-1}\Delta a_i$ for the third and fourth subsystems in (\ref{eq11}), and differentiating $V_i$ with respect to time along the trajectories of these two subsystems and exploiting {Property 2}, we have  $\dot{V}_i=-s_i^T K_i s_i\le 0$, which then yields the result that $s_i\in {\cal L}_2\cap {\cal L}_\infty$ and $\hat{a}_i\in {\cal L}_\infty$, $\forall i=1,2,\dots,n$.

The result that $s_i\in {\cal L}_2\cap {\cal L}_\infty, \forall i=1,2,\dots,n$ implies that $s_E^\ast\in {\cal L}_2\cap {\cal L}_\infty$. From Lemma 2, we know that all the poles of $\Psi_r$ in (\ref{eq18}) are in the open LHP in the case that the graph contains a spanning tree rooted at vertex 0. In addition, it is obvious that the relative degree of the second subsystem in (\ref{eq18}) is two if $\sigma_E$ is taken as the output and $s_E^\ast$ as the input. Therefore, the input-output mapping described by the second subsystem in (\ref{eq18}) is exponentially stable and strictly proper. From the input-output properties of linear systems \cite[p.~59]{Desoer1975_Book}, we obtain that $\sigma_E\in {\cal L}_2\cap {\cal L}_\infty$, $\dot{\sigma}_E\cap {\cal L}_2$, and $\sigma_E\to 0$ as $t\to\infty$. Rewrite the second subsystem in (\ref{eq17}) as
\be
\label{eq19}
\underbrace{\dot{\xi}_E=-\left(\bar{L}_w^\ast\otimes I_m\right)\xi_E}_{\Psi_{rr}}+[\underbrace{-\alpha\sigma_E+s_E^\ast}_\text{system input}]
\ee
where the system input $-\alpha\sigma_E+s_E^\ast\in {\cal L}_2\cap {\cal L}_\infty$, and the system $\Psi_{rr}:\dot{\xi}_E=-\left(\bar{L}_w^\ast\otimes I_m\right)\xi_E$ is obviously exponentially stable and strictly proper if $\xi_E$ is taken as the output signal since, from the similarity decomposition (\ref{eq16}), all the eigenvalues of $\bar{L}_w^\ast$ are in the open RHP (implying that $-\bar{L}_w^\ast$ is Hurwitz) if the graph contains a spanning tree rooted at vertex 0. Therefore, from the input-output properties of linear systems \cite[p.~59]{Desoer1975_Book}, we obtain that $\xi_E\in {\cal L}_2\cap {\cal L}_\infty$, $\dot{\xi}_E\in {\cal L}_2$, and $\xi_E\to 0$ as $t\to\infty$. It is also obvious that $\dot{\xi}_E\in {\cal L}_\infty$ since all the variables on the right side of (\ref{eq19}) is bounded.

From the standard linear system theory, the explicit solution of the first subsystem in (\ref{eq18}) can
be written as
\be
\label{eq25}
\begin{bmatrix}
\int_0^t q_0(r)dr\\
q_0(t)
\end{bmatrix}=\begin{bmatrix}\frac{1}{\sqrt{\alpha}}\sin(\sqrt{\alpha}t) & \frac{1}{\alpha}\left[1-\cos(\sqrt{\alpha}t)\right]\\ \cos(\sqrt{\alpha}t) & \frac{1}{\sqrt{\alpha}}\sin(\sqrt{\alpha}t)\end{bmatrix}\begin{bmatrix}q_0(0)\\ \dot{q}_0(0)\end{bmatrix}
\ee
where it is obvious that $\int_0^t q_0(r)dr\in {\cal L}_\infty$, $q_0(t)\in {\cal L}_\infty$, and $\dot{q}_0(t)=-\sqrt{\alpha}\sin (\sqrt{\alpha}t)q_0(0)+\cos(\sqrt{\alpha}t)\dot{q}_0(0)\in {\cal L}_\infty$, $\forall t\ge 0$. Then, from the result that $\int_0^t \xi_E(r)dr=\sigma_E+\alpha^{-1}d_0\in {\cal L}_\infty$, $\xi_E\in {\cal L}_\infty$, and $\dot\xi_E\in {\cal L}_\infty$, we obtain that $\int_0^t q_i(r)dr\in {\cal L}_\infty$, $q_i\in {\cal L}_\infty$, and $\dot{q}_i\in {\cal L}_\infty$, $\forall i=1,2,\dots,n$. From (\ref{eq4}) and (\ref{eq5}), we obtain that $\dot{q}_{r,i}\in {\cal L}_\infty$ and $\ddot{q}_{r,i}\in {\cal L}_\infty$, $\forall i=1,2,\dots,n$. Based on (\ref{eq10}), we obtain that $\dot{s}_i\in {\cal L}_\infty$ since $M_i(q_i)$ is uniformly positive definite (by {Property 1}), $\forall i=1,2,\dots,n$. From the definition of $\dot{s}_i$, [i.e., equation (\ref{eq7})], we obtain that $\ddot{q}_i\in {\cal L}_\infty$, $\forall i=1,2,\dots,n$. It can also be easily observed that $\ddot{q}_0(t)=-\alpha\cos(\sqrt{\alpha}t)q_0(0)-\sqrt{\alpha}\sin(\sqrt{\alpha}t)\dot{q}_0(0)\in {\cal L}_\infty$, $\forall t\ge 0$. Therefore, $\ddot{\xi}_E\in{\cal L}_\infty$, implying the uniform continuity of $\dot{\xi}_E$. Then, from Barbalat's Lemma \cite{Slotine1991_Book}, we have $\dot{\xi}_E\to0$ as $t\to\infty$. The result that $\xi_E\to0$ and $\dot{\xi}_E\to 0$ as $t\to\infty$ directly gives the conclusion that $q_i(t)\to q_0(t)$ and $\dot{q}_i(t)\to \dot{q}_0(t)$ as $t\to\infty$, $\forall i=1,2,\dots,n$. \hfill {\small $\blacksquare$}

\section{Simulation Results}

In this section, we show the synchronizing performance of the proposed adaptive control scheme by conducting a simulation with a network of a virtual leader (i.e., agent 0) and nine mass agents (the same as the case in \cite{Cheah2009_Aut,Mei2012_Aut}). These ten  agents interact on a directed graph containing a spanning tree (Fig. 1). The mass agents (i.e., the followers) are assumed to move in the X-Y plane and be governed by the following dynamics \cite{Cheah2009_Aut}
\be
m_i \ddot{q}_i+c_i \dot{q}_i=\tau_i
\ee
where $m_i$ and $c_i$ denote the mass and the damping coefficient of the $i$-th agent, respectively, $\tau_i$ is the control input, and $q_i=\left[x_i,y_i\right]^T$ denotes the position of the $i$-th agent, $i=1,2,\dots,9$. The mass parameters of the agents are $m_1=1.0$, $m_2=1.5$, $m_3=1.6$, $m_4=1.2$, $m_5=0.5$, $m_6=2.5$, $m_7=2.2$, $m_8=1.8$, and $m_9=2.1$. The damping coefficients of the agents are $c_1=0.3$, $c_2=0.5$, $c_3=0.7$, $c_4=0.35$, $c_5=0.6$, $c_6=0.8$, $c_7=0.9$, $c_8=0.75$, and $c_9=0.85$. The parameter $\alpha$ is set as $\alpha=1.0$. The sampling period in the following simulation is chosen to be 5 ms.

The entries of the weighted adjacency matrix $\mathcal{W}^\ast$ are chosen according to the rule that $w_{ij}=1.0$ if $j\in\mathcal{N}_i^\ast$, and $w_{ij}=0$ otherwise, $\forall i,j=0,1,\dots,9$. The controller parameters $K_i$ and $\Gamma_i$ are chosen as $K_i=20.0I_2$ and $\Gamma_i=2.0I_2$, respectively, $i=1,2,\dots,9$. The physical parameters of the agents $a_i=\left[m_i,c_i\right]^T$, $i=1,2,\dots,9$ are assumed to be unknown. The initial parameter estimates are chosen as $\hat{a}_i(0)=\left[0,0\right]^T$, $i=1,2,\dots,9$. The initial state of the virtual leader is set as $q_0(0)=\left[2,0\right]^T$ and $\dot{q}_0(0)=\left[0,1\right]^T$, and from (\ref{eq25}), we know that the path of the virtual leader is an ellipse in the X-Y plane centered at $(0,0)$ (i.e., $q_0(t)=\left[2\cos(t),\sin(t)\right]^T$). The initial positions of the followers are set to be $q_1(0)=\left[3,2\right]^T$, $q_2(0)=\left[-3,2\right]^T$, $q_3(0)=\left[-3,-2\right]^T$, $q_4(0)=\left[3,-2\right]^T$, $q_5(0)=\left[3,0\right]^T$, $q_6(0)=\left[-3,0\right]^T$, $q_7(0)=\left[3,3\right]^T$, $q_8(0)=\left[-3,3\right]^T$, and $q_9(0)=\left[-3,-3\right]^T$, and their initial velocities are set to be $\dot{q}_i(0)=0$, $i=1,2,\dots,9$. The simulation results are plotted in Fig. 2 and Fig. 3, which show that the positions of the nine mass agents indeed converge to that of the virtual leader (which is oscillatory).


\begin{figure}
\centering
\begin{minipage}[t]{1.0\linewidth}
\centering
\includegraphics[width=3.5in]{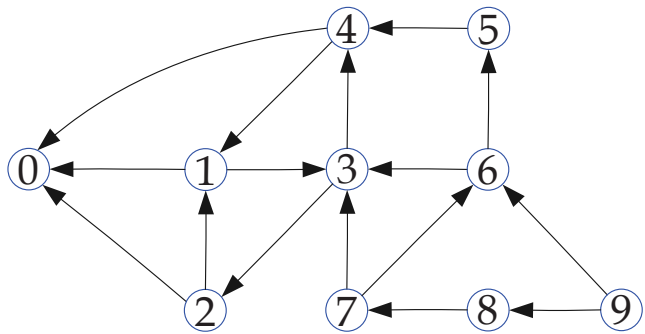}
\caption{Interaction graph among the virtual leader and the 9 followers}\label{fig:side:a}
\end{minipage}%
\end{figure}

%
%


\begin{figure}
\centering
\begin{minipage}[t]{1.0\linewidth}
\centering
\includegraphics[width=3.5in]{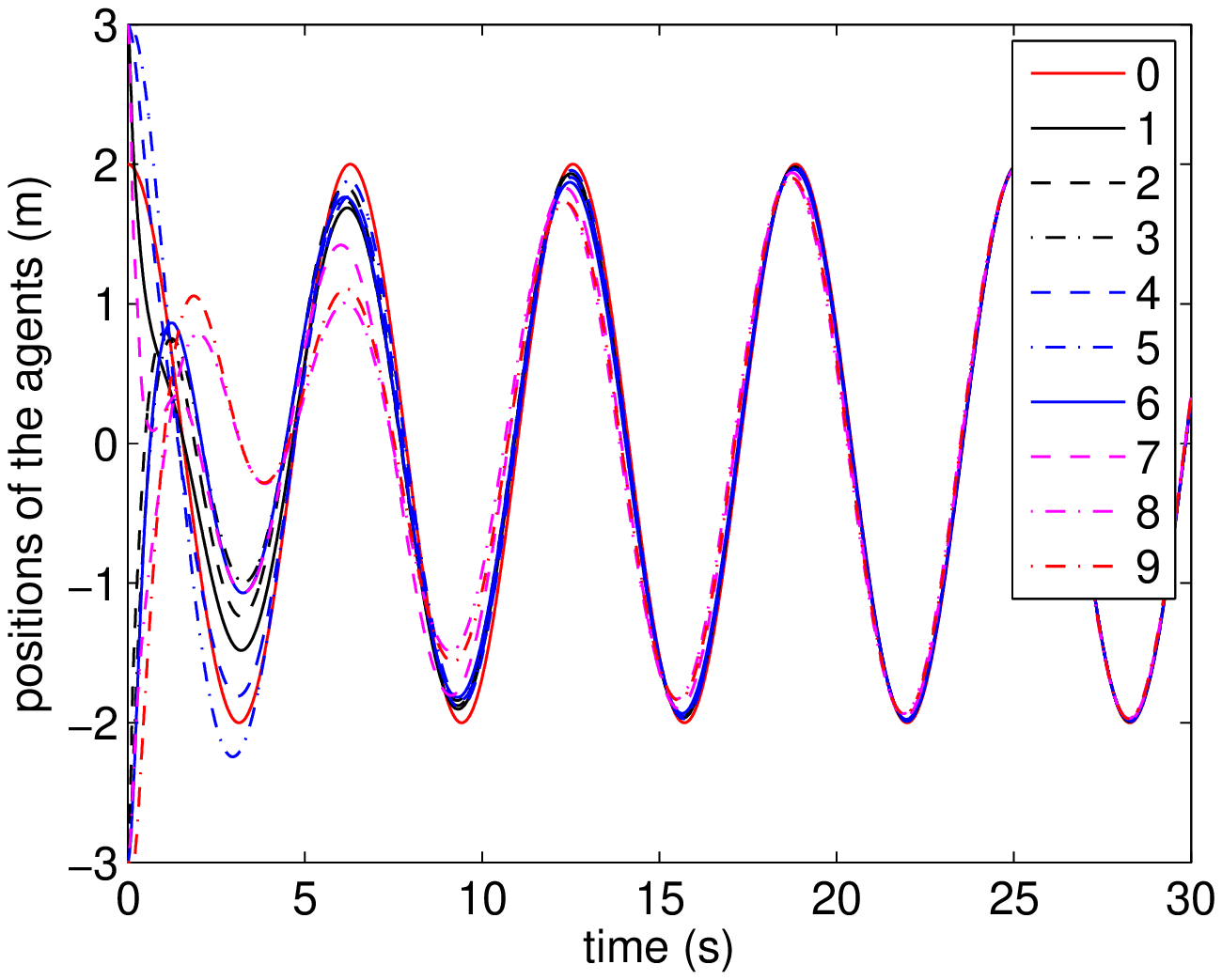}
\caption{Positions of the agents (X-axis)}\label{fig:side:a}
\end{minipage}%
\end{figure}
\begin{figure}
\centering
\begin{minipage}[t]{1.0\linewidth}
\centering
\includegraphics[width=3.5in]{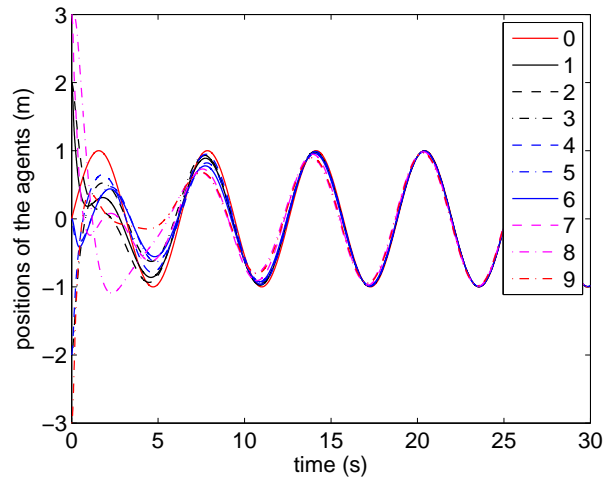}
\caption{Positions of the agents (Y-axis)}\label{fig:side:a}
\end{minipage}%
\end{figure}

\section{Conclusion}

In this paper, we have studied the synchronization problem for a network of multiple mechanical systems (followers) and a virtual leader with oscillatory motion, and the interaction topology among the virtual leader and the mechanical systems is assumed to contain a spanning tree. An adaptive control scheme is proposed to realize the goal of synchronization. Using the similarity decomposition approach, we show that the position and velocity synchronization errors between each mechanical system and the virtual leader converge to zero. A numerical simulation is conducted to illustrate the synchronizing performance of the proposed control scheme.

\section*{Acknowledgement}

The author would like to thank Prof. Wei Ren for the helpful discussions on this topic.

\bibliographystyle{IEEETran}        
\bibliography{..//Reference_list_Wang}           

\begin{thebibliography}{10}
\providecommand{\url}[1]{#1}
\csname url@samestyle\endcsname
\providecommand{\newblock}{\relax}
\providecommand{\bibinfo}[2]{#2}
\providecommand{\BIBentrySTDinterwordspacing}{\spaceskip=0pt\relax}
\providecommand{\BIBentryALTinterwordstretchfactor}{4}
\providecommand{\BIBentryALTinterwordspacing}{\spaceskip=\fontdimen2\font plus
\BIBentryALTinterwordstretchfactor\fontdimen3\font minus
  \fontdimen4\font\relax}
\providecommand{\BIBforeignlanguage}[2]{{%
\expandafter\ifx\csname l@#1\endcsname\relax
\typeout{** WARNING: IEEEtran.bst: No hyphenation pattern has been}%
\typeout{** loaded for the language `#1'. Using the pattern for}%
\typeout{** the default language instead.}%
\else
\language=\csname l@#1\endcsname
\fi
#2}}
\providecommand{\BIBdecl}{\relax}
\BIBdecl

\bibitem{Jadbabaie2003_TAC}
A.~Jadbabaie, J.~Lin, and A.~S. Morse, ``Coordination of groups of mobile
  autonomous agents using nearest neighbor rules,'' \emph{IEEE Transactions on
  Automatic Control}, vol.~48, no.~6, pp. 988--1001, Jun. 2003.

\bibitem{Olfati-Saber2004_TAC}
R.~Olfati-Saber and R.~M. Murray, ``Consensus problems in networks of agents
  with switching topology and time-delays,'' \emph{IEEE Transactions on
  Automatic Control}, vol.~49, no.~9, pp. 1520--1533, Sep. 2004.

\bibitem{Ren2005_TAC}
W.~Ren and R.~W. Beard, ``Consensus seeking in multiagent systems under
  dynamically changing interaction topologies,'' \emph{IEEE Transactions on
  Automatic Control}, vol.~50, no.~5, pp. 655--661, May 2005.

\bibitem{Jadbabaie2004_ACC}
A.~Jadbabaie, N.~Motee, and M.~Barahona, ``On the stability of the {K}uramoto
  model of coupled nonlinear oscillators,'' in \emph{Proceedings of the
  American Control Conference}, Boston, MA, 2004, pp. 4296--4301.

\bibitem{Chopra2006}
N.~Chopra and M.~W. Spong, ``Passivity-based control of multi-agent systems,''
  in \emph{Advances in Robot Control: From Everyday Physics to Human-Like
  Movements}, S.~Kawamura and M.~Svinin, Eds.\hskip 1em plus 0.5em minus
  0.4em\relax Berlin: Springer-Verlag, 2006, pp. 107--134.

\bibitem{Ren2007_RNC}
W.~Ren and E.~Atkins, ``Distributed multi-vehicle coordinated control via local
  information exchange,'' \emph{International Journal of Robust and Nonlinear
  Control}, vol.~17, no. 10-11, pp. 1002--1033, Jul. 2007.

\bibitem{Lee2007_TAC}
D.~Lee and M.~W. Spong, ``Stable flocking of multiple inertial agents on
  balanced graphs,'' \emph{IEEE Transactions on Automatic Control}, vol.~52,
  no.~8, pp. 1469--1475, Aug. 2007.

\bibitem{Ren2008_Aut}
W.~Ren, ``Synchronization of coupled harmonic oscillators with local
  interaction,'' \emph{Automatica}, vol.~44, no.~12, pp. 3195--3200, Dec. 2008.

\bibitem{Chopra2009_TAC}
N.~Chopra and M.~W. Spong, ``On exponential synchronization of {K}uramoto
  oscillators,'' \emph{IEEE Transactions on Automatic Control}, vol.~54, no.~2,
  pp. 353--357, Feb. 2009.

\bibitem{Su2009_Aut}
H.~Su, X.~Wang, and Z.~Lin, ``Synchronization of coupled harmonic oscillators
  in a dynamic proximity network,'' \emph{Automatica}, vol.~45, no.~10, pp.
  2286--2291, Oct. 2009.

\bibitem{Cheah2009_Aut}
C.~C. Cheah, S.~P. Hou, and J.-J.~E. Slotine, ``Region-based shape control for
  a swarm of robots,'' \emph{Automatica}, vol.~45, no.~10, pp. 2406--2411, Oct.
  2009.

\bibitem{Wang2013cyclic_ASME}
H.~Wang and Y.~Xie, ``Cyclic constraint analysis for attitude synchronization
  of networked spacecraft agents,'' \emph{Journal of Dynamic Systems,
  Measurement, and Control}, vol. 135, no.~6, pp. 061\,019--1--061\,019--8,
  Nov. 2013.

\bibitem{Ren2008_Book}
W.~Ren and R.~W. Beard, \emph{Distributed Consensus in Multi-Vehicle
  Cooperative Control}.\hskip 1em plus 0.5em minus 0.4em\relax London:
  Springer-Verlag, 2008.

\bibitem{Nuno2011_TAC}
E.~Nu{\~n}o, R.~Ortega, L.~Basa{\~n}ez, and D.~Hill, ``Synchronization of
  networks of nonidentical {E}uler-{L}agrange systems with uncertain parameters
  and communication delays,'' \emph{IEEE Transactions on Automatic Control},
  vol.~56, no.~4, pp. 935--941, Apr. 2011.

\bibitem{Wang2013b_Aut}
H.~Wang, ``Flocking of networked uncertain {E}uler-{L}agrange systems on
  directed graphs,'' \emph{Automatica}, vol.~49, no.~9, pp. 2774--2779, Sep.
  2013.

\bibitem{Wang2013_TAC}
------, ``Task-space synchronization of networked robotic systems with
  uncertain kinematics and dynamics,'' \emph{IEEE Transactions on Automatic
  Control}, vol.~58, no.~12, pp. 3169--3174, Dec. 2013.

\bibitem{Paley2007_CSM}
D.~A. Paley, N.~E. Leonard, R.~Sepulchre, D.~Grunbaum, and J.~K. Parrish,
  ``Oscillator models and collective motion,'' \emph{IEEE Control Systems
  Magazine}, vol.~27, no.~4, pp. 89--105, Aug. 2007.

\bibitem{Yu2010_SMC}
W.~Yu, G.~Chen, M.~Cao, and J.~Kurths, ``Second-order consensus for multiagent
  systems with directed topologies and nonlinear dynamics,'' \emph{IEEE
  Transactions on Systems, Man, and Cybernetics-Part B: Cybernetics}, vol.~40,
  no.~3, pp. 881--891, Jun. 2010.

\bibitem{Liu2013_SCL}
K.~Liu, G.~Xie, W.~Ren, and L.~Wang, ``Consensus for multi-agent systems with
  inherent nonlinear dynamics under directed topologies,'' \emph{Systems \&
  Control Letters}, vol.~62, no.~2, pp. 152--162, Feb. 2013.

\bibitem{Su2011_Aut}
H.~Su, G.~Chen, X.~Wang, and Z.~Lin, ``Adaptive second-order consensus of
  networked mobile agents with nonlinear dynamics,'' \emph{Automatica},
  vol.~47, no.~2, pp. 368--375, Feb. 2011.

\bibitem{Li2013_TAC}
Z.~Li, W.~Ren, X.~Liu, and M.~Fu, ``Consensus of multi-agent systems with
  general linear and {L}ipschitz nonlinear dynamics using distributed adaptive
  protocols,'' \emph{IEEE Transactions on Automatic Control}, vol.~58, no.~7,
  pp. 1786--1791, Jul. 2013.

\bibitem{Slotine1991_Book}
J.-J.~E. Slotine and W.~Li, \emph{Applied Nonlinear Control}.\hskip 1em plus
  0.5em minus 0.4em\relax Englewood Cliffs, NJ: Prentice-Hall, 1991.

\bibitem{Mei2012_Aut}
J.~Mei, W.~Ren, and G.~Ma, ``Distributed containment control for {L}agrangian
  networks with parametric uncertainties under a directed graph,''
  \emph{Automatica}, vol.~48, no.~4, pp. 653--659, Apr. 2012.

\bibitem{Cai2013_CCC}
H.~Cai and J.~Huang, ``Distributed leader-following consensus for multiple
  {E}uler-{L}agrange systems under switching network,'' in \emph{Proceedings of
  the Chinese Control Conference}, Xi'an, China, 2013, pp. 7228--7233.

\bibitem{Ding2013_TAC}
Z.~Ding, ``Consensus output regulation of a class of heterogeneous nonlinear
  systems,'' \emph{IEEE Transactions on Automatic Control}, vol.~58, no.~10,
  pp. 2648--2653, Oct. 2013.

\bibitem{Ding2013_CDC}
------, ``Adaptive consensus output regulation of a class of heterogeneous
  nonlinear systems,'' in \emph{Proceedings of the IEEE Conference on Decision
  and Control}, Florence, Italy, 2013, pp. 5385--5390.

\bibitem{Min2012_SCL}
H.~Min, S.~Wang, F.~Sun, Z.~Gao, and J.~Zhang, ``Decentralized adaptive
  attitude synchronization of spacecraft formation,'' \emph{Systems \& Control
  Letters}, vol.~61, no.~1, pp. 238--246, Jan. 2012.

\bibitem{Wang2014_TAC}
H.~Wang, ``Consensus of networked mechanical systems with communication delays:
  {A} unified framework,'' \emph{IEEE Transactions on Automatic Control},
  vol.~59, no.~6, pp. 1571--1576, Jun. 2014.

\bibitem{Godsil2001_Book}
C.~Godsil and G.~Royle, \emph{Algebraic Graph Theory}.\hskip 1em plus 0.5em
  minus 0.4em\relax New York: Springer-Verlag, 2001.

\bibitem{Spong2006_Book}
M.~W. Spong, S.~Hutchinson, and M.~Vidyasagar, \emph{Robot Modeling and
  Control}.\hskip 1em plus 0.5em minus 0.4em\relax New York: John Wiley \&
  Sons, 2006.

\bibitem{Slotine1987_IJRR}
J.-J.~E. Slotine and W.~Li, ``On the adaptive control of robot manipulators,''
  \emph{The International Journal of Robotics Research}, vol.~6, no.~3, pp.
  49--59, Sep. 1987.

\bibitem{Brewer1978_TCS}
J.~W. Brewer, ``Kronecker products and matrix caculus in system theory,''
  \emph{IEEE Transactions on Circuits and Systems}, vol. CAS-25, no.~9, pp.
  772--781, Sep. 1978.

\bibitem{Li2010_TAC}
W.~Li and M.~W. Spong, ``Stability of general coupled inertial agents,''
  \emph{IEEE Transactions on Automatic Control}, vol.~55, no.~6, pp.
  1411--1416, Jun. 2010.

\bibitem{Ren2005}
W.~Ren, R.~W. Beard, and T.~W. McLain, ``Coordination variables and consensus
  building in multiple vehicle systems,'' in \emph{Cooperative Control}, ser.
  Lecture Notes in Control and Information Science, V.~Kumar, N.~Leonard, and
  A.~S. Morse, Eds.\hskip 1em plus 0.5em minus 0.4em\relax Berlin, Germany:
  Springer-Verlag, 2005, vol. 309, pp. 171--188.

\bibitem{Meng2011_SMC}
Z.~Meng, W.~Ren, Y.~Cao, and Z.~You, ``Leaderless and leader-following
  consensus with communication and input delays under a directed network
  topology,'' \emph{IEEE Transactions on Systems, Man, and Cybernetics-Part B:
  Cybernetics}, vol.~41, no.~1, pp. 75--88, Feb. 2011.

\bibitem{Lee2012_Aut}
D.~Lee, ``Distributed backstepping control of multiple thrust-propelled
  vehicles on a balanced graph,'' \emph{Automatica}, vol.~48, no.~11, pp.
  2971--2977, Nov. 2012.

\bibitem{Ortega1989_Aut}
R.~Ortega and M.~W. Spong, ``Adaptive motion control of rigid robots: {A}
  tutorial,'' \emph{Automatica}, vol.~25, no.~6, pp. 877--888, Nov. 1989.

\bibitem{Desoer1975_Book}
C.~A. Desoer and M.~Vidyasagar, \emph{Feedback Systems: Input-Output
  Properties}.\hskip 1em plus 0.5em minus 0.4em\relax New York: Academic Press,
  1975.

\end{thebibliography}
\end{document}